\documentclass[11pt,letterpaper]{JHEP3}
\usepackage{graphics}
\usepackage{epsfig}
\usepackage{subfigure}
\usepackage{mathrsfs}
\usepackage{amsmath}
\usepackage{amsfonts}
\usepackage{amssymb}
\usepackage{graphicx}

\title{Holographic thermalization in Gauss-Bonnet gravity}
\author{Xiao-Xiong Zeng\\
School of Science, Chongqing Jiaotong University, Chongqing, 400074, China\\
 \email{xxzeng@mail.bnu.edu.cn}}
\author{Wen-Biao Liu\\
Department of Physics, Beijing Normal University, Beijing, 100875, China\\
\email{wbliu@bnu.edu.cn}
}
 \abstract{In the spirit of  AdS/CFT correspondence, we study the thermalization of
a dual conformal field theory to  Gauss-Bonnet gravity by modeling a thin-shell of dust that interpolates between a pure AdS and a Gauss-Bonnet AdS black brane. The  renormalized geodesic length and minimal  area surface, which in the dual conformal  field   theory  correspond to
two-point correlation function and expectation value of Wilson loop, are investigated respectively as thermalization probes. The result shows that as the
 Gauss-Bonnet coefficient increases, the thermalization time decreases for both the thermalization probes, which can also be confirmed by studying the motion profile of the  geodesic and minimal  area surface. In addition, for both the  renormalized geodesic length and minimal  area surface,
  there is an overlapped region for a fixed boundary separation, which implies that the Gauss-Bonnet coupling constant has little effect on the thermalization probes there.

}

\preprint{}
\begin{document}
\bibliographystyle{ieeetr}
\bibliography{reference.bib}
\section{Introduction}
Non-equilibrium phenomena are ubiquitous. In particular, it is now possible to control some systems out of equilibrium in the laboratory due to
various experimental breakthroughs in atomic physics, quantum optics and nanoscience. With this, a wealth of effort has been made towards the theoretical understanding of non-equilibrium physics. But nevertheless some non-equilibrium experiments are conducted with strongly coupled systems, which includes quark
gluon plasma produced in RHIC and LHC experiments and cold atomic gases prepared in quantum quenches. Faced up with the non-equilibrium physics for the strongly coupled systems, the available approaches are limited, where the string inspired AdS/CFT correspondence stands out as an unconventional but suitable tool. As sort of strong weak
duality, AdS/CFT correspondence maps the problem of strongly coupled systems to a much easier bulk dynamics with one extra dimension, where the static black hole in the bulk is dual to the boundary
system in equilibrium with finite temperature, and the small perturbation on top of the black hole drives the boundary system to
a near-equilibrium state. Since its advent, such a paradigm has provided us with various remarkable insights into our understanding of universal equilibrium and its linear response
behaviors of strongly coupled systems. However, compared with this, much less is known about the universal far-from-equilibrium
behaviors of strongly coupled systems because the corresponding holographic bulk is supposed to be a highly dynamical spacetime involving black hole formation or black hole merger, where the involved numerical techniques are also non-trivial due to the non-linearity of the bulk differential equation.

One way out is to play with some toy models, which may be too simple to model the realistic systems but subject to analytic control
such as to provide us with some insights into the universal aspects of far-from-equilibrium physics for strongly coupled systems\cite{Garfinkle84, Garfnkle1202, Allais1201, Das343, Steineder, Wu1210, Gao, Li}. Among
others, very recently the authors in \cite{Balasubramanian1} as well as \cite{Balasubramanian2} have used a neutral AdS-Vaidya black hole
as the bulk geometry to probe the scale-dependence of holographic thermalization following a quench via calculations of two-point
correlation functions, Wilson loops, and entanglement entropy, which can further be evaluated in the saddle point approximation in terms
of geodesics, minimal surfaces, and minimal volume individually. It is found that the holographic thermalization always proceeds in a
top-down pattern, namely the UV modes thermalize, followed by the IR modes. They also find that there is a slight delay in the onset of thermalization and the entanglement entropy thermalizes slowest, which sets a timescale for equilibration.  Later, such an investigation is generalized to the bulk
geometry given by a charged AdS-Vaidya black hole to see how the chemical potential affects the holographic thermalization, as by
holography the charged black hole corresponds to the boundary system with a finite chemical potential\cite{GS,CK}.  Recently there are more and more extensions \cite{Baron, Baron1212, Arefeva, Hubeny, Balasubramanianeyal6066, Arefeva6041} to study holographic thermalization on the basis of the work in   \cite{Balasubramanian1, Balasubramanian2}.

The purpose of this paper is to investigate how the holographic thermalization behaves in Gauss-Bonnet gravity. By holography, Gauss-Bonnet gravity corresponds to $\frac{1}{N}$ or $\frac{1}{\lambda}$ correction to the boundary field theory, depending on whether the origin of Gauss-Bonnet term comes from stringy or quantum effect.  In light of this reason,  there have been many works
to study the strong coupling system  of
the dual conformal field theory in the Gauss-Bonnet gravity to explore
the
effects of  Gauss-Bonnet coefficient on the observables \cite{Gregory010, Pan, cai066027, Wu106, Hu123, 0806.1334}.
 As to holographic thermalization in Gauss-Bonnet gravity, we will take the
two-point function and expectation value of Wilson loops as thermalization probes to study the
thermalization behavior. According to  the AdS/CFT correspondence, this process equals to probe the evolution of a shell of
 dust that interpolates between a pure AdS and a Gauss-Bonnet AdS black brane by making use of the
   renormalized geodesic lengths and minimal  area surfaces.
 Concretely we first study the motion profile of the geodesic and minimal area, and then the  renormalized geodesic length and minimal  area surface
 in the  Gauss-Bonnet  Vaidya AdS black brane, the result shows that  larger the Gauss-Bonnet coefficient is,  easier the dual boundary system
 thermalizes.  We also study the thermalization time for both the thermalization probes at different boundary separation, and the result
 shows that the  UV modes thermalize first, which is the same as the situation occurs in the Einstein gravity. In addition, we find that
  there is an  overlapped region for both the thermalization probes at a fixed boundary separation,  where the Gauss-Bonnet coefficient has little effect on the  renormalized geodesic length and minimal  area surface. We also analyze the reason that leads to this phenomenon.

The remainder of this paper is structured as follows. In the next section, we shall provide a brief review of Vaidya AdS black brane in Gauss-Bonnet gravity. Then
the holographic setup for non-local observables will be explicitly constructed in Section 3. Resorting to numerical calculation, we
shall perform a systematic analysis of how the Gauss-Bonnet coefficient affects the thermalization time in Section 4. We end
up with some discussions in the last section.

\section{Vaidya AdS black branes in Gauss-Bonnet gravity}
Start from the
action of the $D(D \geq 5)$ dimensional Gauss-Bonnet gravity with a negative
cosmological constant
\begin{equation}
\label{action} I=\frac{1}{16 \pi G}\int_\mathcal{M}d^Dx
\sqrt{-g} \left(R-2 \Lambda+\alpha L_{GB} \right),
\end{equation}
where $G$ is the D-dimensional gravitational constant, $R$ is the Ricci scalar, $\Lambda$ is the negative cosmological constant, and $\alpha$
is the Gauss-Bonnet coefficient with the Gauss-Bonnet term $L_{GB}$
given by
\begin{equation}
\label{LGB} L_{GB} =R^2-4R_{\mu \nu}R^{\mu \nu}+R_{\mu \nu \sigma
\tau}R^{\mu \nu \sigma \tau}.
\end{equation}
Through the variation of the action (\ref{action}) with respect to
the bulk metric, one can obtain the equation of motion for Gauss-Bonnet
gravity, i.e.,
\begin{equation}
R_{\mu \nu } -\frac{1}{2}Rg_{\mu \nu}+\Lambda g_{\mu \nu
}+\alpha H_{\mu \nu}=0,
\end{equation}
where
\begin{equation}
H_{\mu \nu}=2(R_{\mu \sigma \kappa \tau }R_{\nu }^{\phantom{\nu}
\sigma \kappa \tau }-2R_{\mu \rho \nu \sigma }R^{\rho \sigma
}-2R_{\mu
\sigma }R_{\phantom{\sigma}\nu }^{\sigma }+RR_{\mu \nu })-\frac{1}{2}
L_{GB}g_{\mu \nu }.
\end{equation}
Accordingly the black brane solution can be obtained as \cite{Cai}
\begin{equation}
ds^{2}=-H(r)dt^{2}+H^{-1}(r)dr^{2}+\frac{r^{2}}{\ell^{2}}d\mathbf{x}^{2},
\end{equation}
where
\begin{equation}\label{blackening}
H(r)=\frac{r^{2}}{2\tilde{\alpha}}\left[1-\sqrt{1-\frac{4\tilde{\alpha}}{\ell^{2}}\left(1-\frac{M
\ell^{2}}{r^{D-1}}\right)}\right],
\end{equation}
with $M$ the mass parameter, $\tilde{\alpha}=(D-3)(D-4)\alpha$ and
$\ell^2 = - \frac{(D-1) (D-2)}{2\Lambda}$. By the regularity of conic singularity in the Euclidean sector, the Hawking temperature, which is also the temperature of the dual conformal field theory,
 is given by
\begin{equation}\label{temperature}
T=\frac{\partial_rH(r)}{4\pi}|_{r_h}=\frac{D-1}{4\pi}M^\frac{1}{D-1}\ell^\frac{4-2D}{D-1},
\end{equation}
where the horizon $r_h=(M\ell^2)^\frac{1}{D-1}$. On the other hand,
As $r$  approaches to infinity, one can see the above black brane metric changes into
\begin{equation}
ds^2\rightarrow \frac{r^2}{\ell^2_{eff}}(-dt^2+d\mathbf{\tilde{x}}^2)+\frac{\ell^2_{eff}}{r^2}dr^2,
\end{equation}
where
\begin{equation}
\mathbf{\tilde{x}}=\frac{\ell_{eff}}{\ell}\mathbf{x},
\end{equation}
and
\begin{equation}
\ell^2_{eff}=\frac{2\tilde{\alpha}}{1-\sqrt{1-\frac{4\tilde{\alpha}}{\ell^{2}}}}.
\end{equation}
Thus this black brane solution is asymptotically AdS with AdS radius $\ell_{eff}$.

From Eq.(\ref{blackening}), one can see  that there is an upper bound for the
Gauss-Bonnet coefficient, namely $\tilde{\alpha}\leq \ell^2/4$. This is known as the
Chern-Simons limit. Besides there also exists a constraint
$-\frac{(3D-1)(D-3)}{4(D+1)^{2}}\ell^2\leq\tilde{\alpha}\leq\frac{(D-3)(D-4)(D^2-3D+8)}{4(D^2-5D+10)^2}\ell^2$ by demanding the
causality of dual field theory on the boundary\cite{0911.3160, 0912.1944, Buchel}.

To get a Vaidya type evolving black brane, we would like first to make the coordinate
transformation $z=\frac{\ell^2}{r}$, with which the above black brane metric can be cast into
\begin{equation}
ds^{2}=\frac{\ell^2}{z^2}[-H(z)dt^{2}+H^{-1}(z)dz^{2}+d\textbf{x}^2],
\end{equation}
where
\begin{equation}
H(z)=\frac{\ell^2}{2\tilde{\alpha}}\left[1-\sqrt{1-\frac{4\tilde{\alpha}}{\ell^2}\left(1-M
z^{D-1}\ell^{4-2D}\right)}\right].
\end{equation}
 Then by introducing the
Eddington-Finkelstein coordinate system, namely
\begin{equation}
dv=dt-\frac{1}{H(z)}dz,
\end{equation}
one can obtain
\begin{equation}
ds^2=\frac{\ell^2}{z^2} \left[ - H(z) d{v}^2 - 2 dz\ dv +
d\textbf{x}^2 \right] .
\end{equation}
Now the Gauss-Bonnet Vaidya AdS  black brane\footnote{This solution is also obtained in \cite{Dominguez73}
by adding the action in (\ref{action}) with a matter field $S_{matter}$. } can be obtained by
freeing the mass parameter as an arbitrary function of $v$\cite{Kobayashi,Maeda}. As one can show,
such a metric is sourced by the null dust with the energy momentum tensor as
\begin{equation}
T_{\mu\nu}\propto(D-2)z^{D-2}\frac{dM(v)}{dv}\delta_{\mu v}\delta_{\nu v}.
\end{equation}
According to the AdS/CFT correspondence, the rapid injection of energy followed by the thermalization process on the boundary corresponds to the collapse of a black
brane in the AdS space.
So to describe the
thermalization process holographically, one should choose the mass
as the function of time $v$ so that in the limit $v\rightarrow
-\infty$, the background  corresponds to a pure AdS space while in
the limit $v\rightarrow \infty$,
 it corresponds to a Gauss-Bonnet AdS black brane, which can be definitely  achieved by setting
the mass parameter $M(v)=M\theta({v})$ with $\theta({v})$ the step function.
But for the convenience of later numerical calculations, $M(v)$ is usually
chosen as a smooth function
\begin{equation}
M(v) = \frac{M}{2} \left( 1 + \tanh \frac{v}{v_0} \right),
\end{equation}
where $v_0$ represents a finite shell thickness.

For simplicity but without loss of generality, we shall set  the unit $\ell=1$ in the later discussions. In addition, $M$ is also set to one as the situation for other magnitudes of the mass parameter can be readily obtained by rescaling the coordinates.

\section{Holographic setup for non-local observables}
Having the construction of a model that describes the thermalization
process on the dual conformal field theory, we have to choose a set of extended observables in the bulk
which allow us to evaluate the evolution of the system.
 For simplicity, here we shall focus mainly on the two-point correlation
function at equal time and expectation value of rectangular space-like Wilson loop.
\subsection{Two-point correlation function at equal time}
According to the AdS/CFT correspondence,  the equal time two-point correlation function under  the saddle-point
approximation can
be holographically approximated as \cite{Balasubramanian2, Balasubramanian61}
\begin{equation}
\langle {\cal{O}} (t_0,\textbf{x}) {\cal{O}}(t_0, \textbf{x}')\rangle  \approx
e^{-\Delta {\tilde{L}_{ren}}} ,\label{llll}
\end{equation}
if the conformal dimension $\Delta$ of scalar operator $\cal{O}$
is large enough, where
$\tilde{L}_{ren}$ indicates the renormalized length of the bulk geodesic between the points $(t_0,
\textbf{x})$ and $(t_0, \textbf{x}')$ on the AdS boundary.
 Based on Eq.(\ref{llll}), we will concentrate on studying the space-like geodesic in the  Gauss-Bonnet gravity to explore how the  Gauss-Bonnet coefficient affects the thermalization time.

For the AdS black brane in Gauss-Bonnet gravity, it is asymptotically AdS with AdS radius $\ell_{eff}$
 and boundary coordinate $\mathbf{\tilde{x}}$.  Taking into account the spacetime symmetry of our Vaidya type black brane,
  we can simply let $\mathbf{x}$ and $\mathbf{x'}$ have identical
  coordinates except $\tilde{x}^1=-\ell_{eff}\frac{l}{2}$ and $\tilde{x}'^1=\ell_{eff}\frac{l}{2}$
  with $\ell_{eff} l$ the separation between these two points on the boundary, where $l$ is the boundary
  separation of the  Vaidya black brane  as discussed in \cite{Balasubramanian1,Balasubramanian2}. In order to make
   the notation as simple as possible, we would like to rename this exceptional coordinate $x^1$ as $x$ and employ it to
    parameterize the trajectory such that the proper length is given by

\begin{equation}
  \tilde{L}  =  \int_{-\ell_{eff}\frac{l}{2}}^{\ell_{eff}\frac{l}{2}} dx
\frac{\sqrt{1-2z'(x)v'(x) - H(v,z) v'(x)^2}}{z(x)} ,\label{false}
\end{equation}
where the prime denotes the derivative with respect to $x$ and
\begin{equation}
H(v,z)=\frac{1}{2\tilde{\alpha}}\left[1-\sqrt{1-4\tilde{\alpha}\left(1-M(v)
z^{D-1}\right)}\right].
\end{equation}
Note that the integrand in
Eq.(\ref{false}) can be thought of as the Lagrangian $\cal{L}$ of a fictitious system with $x$ the proper time.
Since the Lagrangian does not depend explicitly on $x$, there is an associated conserved quantity
\begin{equation}
 {\cal{H}}  ={\cal{L}}-v'(x)\frac{\partial \cal{L}}{\partial v'(x) } -  z'(x)\frac{\partial \cal{L}}{\partial z'(x) }= \frac{1}{z(x) \sqrt{1-2z'(x)v'(x) - H (v,z) v'(x)^2} }.
\end{equation}
With it, the equations of motion for $z(x)$ and $v(x)$ can be obtained as
\begin{eqnarray} \label{gequation}
0 &=& 2 - 2 v'(x)^2 H(v,z) - 4 v'(x) z'(x) - 2 z(x) v''(x) + z(x)
v'(x)^2 \partial_z H(v,z),\nonumber\\
0&=&v'(x)z'(x)\partial_zH(v,z)+\frac{1}{2}v'(x)^2\partial_vH(v,z)+v''(x)H(v,z)+z''(x).
\end{eqnarray}
Furthermore, by the reflection symmetry of our geodesic, we have the following initial conditions
\begin{equation}\label{initial}
z(0)=z_*,  v(0)=v_* , v'(0) =
z'(0) = 0.
\end{equation}
Thus the proper length of geodesic in (\ref{false}) can be simplified as
\begin{equation}
\tilde{L} = 2 \int_{0}^{\ell_{eff}\frac{l}{2}} dx \frac{z_*}{z(x)^2} .
\end{equation}
Generically this proper length is divergent. So one needs to make regularization and renormalization. The regularization can be achieved by
imposing the boundary conditions as follows
\begin{equation}\label{regularization}
z(\ell_{eff}\frac{l}{2})=z_0, v(\ell_{eff}\frac{l}{2})=t_0,
\end{equation}
where $z_0$ is the IR radial cut-off. Then by subtracting the divergent part\footnote{This part is the contribution of the geodesic length near the AdS boundary, one can refer \cite{CK} to get the details.}, one ends up with the renormalized geodesic length as
\begin{equation}\label{lren}
\tilde{L}_{ren}=2 \int_{0}^{\ell_{eff}\frac{l}{2}} dx \frac{z_*}{z(x)^2}+2\ell_{eff}\ln z_0.
\end{equation}

\subsection{Expectation value of rectangular space-like Wilson loop}
Wilson loop operator is defined as a path ordered integral of gauge field over a closed contour,
 and its expectation value is approximated geometrically  by the AdS/CFT correspondence as \cite{Balasubramanian2, Maldacena80}
\begin{equation}
\langle W(C)\rangle \approx e^{-\frac{\tilde{A}_{ren}(\Sigma)}{2\pi\alpha'}},
\end{equation}
where $C$ is the closed contour, $\Sigma$ is the minimal bulk surface ending on $C$ with $\tilde{A}_{ren}$ its renormalized
minimal  area surface  , and  $\alpha'$ is the Regge slope parameter.

Here we are focusing solely on the rectangular space-like Wilson loop. In this case, the enclosed rectangle can be always
chosen to be centered at the coordinate origin and lying on the $x^1-x^2$ plane with the assumption that the corresponding
 bulk surface is invariant along the $x^2$ direction. This implies that the minimal  area surface   can be expressed as
\begin{equation}
\tilde{A}=\int_{-\ell_{eff}\frac{l}{2}}^{\ell_{eff}\frac{l}{2}}dx
\frac{\sqrt{1-2z'(x)v'(x) - H(v,z) v'(x)^2}}{z(x)^2},
\end{equation}
where we have set the separation along $x^2$ direction to be one and the separation along $x^1$ to be $\ell_{eff}l$ with $x^2$ renamed as $y$ and $x^1$ renamed as $x$.

As before, we have also a conserved quantity, i.e.,
\begin{equation}
{\cal{H}}=\frac{1}{z(x)^2 \sqrt{1-2z'(x)v'(x) - H (v,z) v'(x)^2} },
\end{equation}
which can simplify our equations of motion as
\begin{eqnarray} \label{aequation}
0&=&4-4v'(x)^2H(v,z)-8v'(x)z'(x)-2z(x)v''(x)+z(x)v'(x)^2\partial_zH(v,z),\nonumber\\
0&=&v'(x)z'(x)\partial_zH(v,z)+\frac{1}{2}v'(x)^2\partial_vH(v,z)+v''(x)H(v,z)+z''(x).
\end{eqnarray}
Similarly, with the initial conditions as in (\ref{initial}) and the regularization cut-off as in (\ref{regularization}), the
renormalized minimal  area surface  can be cast into
\begin{equation}\label{aren}
\tilde{A}_{ren}=2\int_0^{\ell_{eff}\frac{l}{2}}dx
\frac{z^2_*}{z(x)^4}-\frac{2}{z_0}.
\end{equation}

\section{Numerical results}
In this section,  we concentrate on finding the renormalized geodesic length and
minimal  area surface  numerically on the basis of  (\ref{lren}) and  (\ref{aren}). Because there have been many works to study the effect of
the space time dimensions on the thermalization probes \cite{Balasubramanian1, Balasubramanian2, GS, CK}, to avoid redundancy,
we mainly discuss the case $D=5$ in this paper. During the numerics, we will take the shell thickness, horizon and  UV
 cut-off as $v_0 = 0.01$, $r_h = 1$, $z_0 = 0.01$ respectively and relabel the boundary separation  $\ell_{eff} l$ as  $\tilde{\ell}$.

\begin{figure}
\centering
\subfigure[$v_*=-0.856, \alpha=-0.1$]{
\includegraphics[scale=0.55]{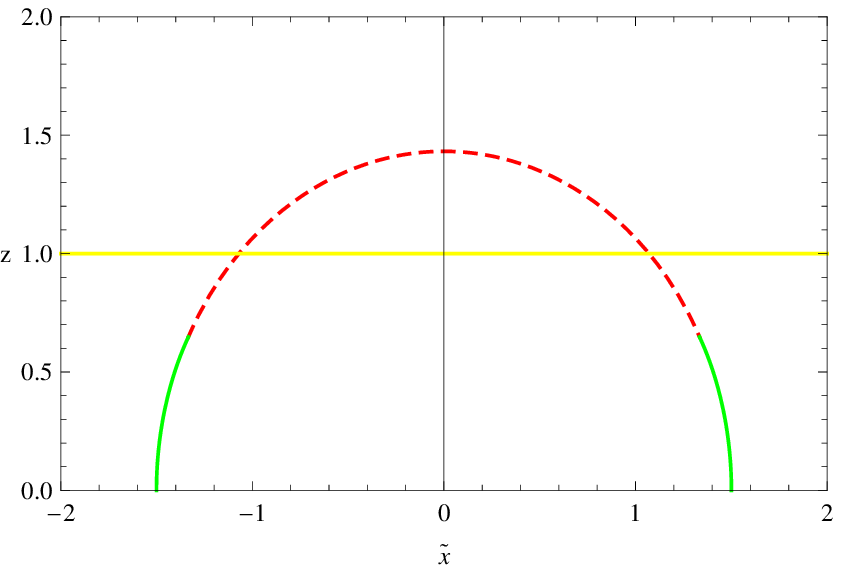}
\label{shell1_RN} }
\subfigure[$v_*=-0.856, \alpha=0.0001$]{
\includegraphics[scale=0.55]{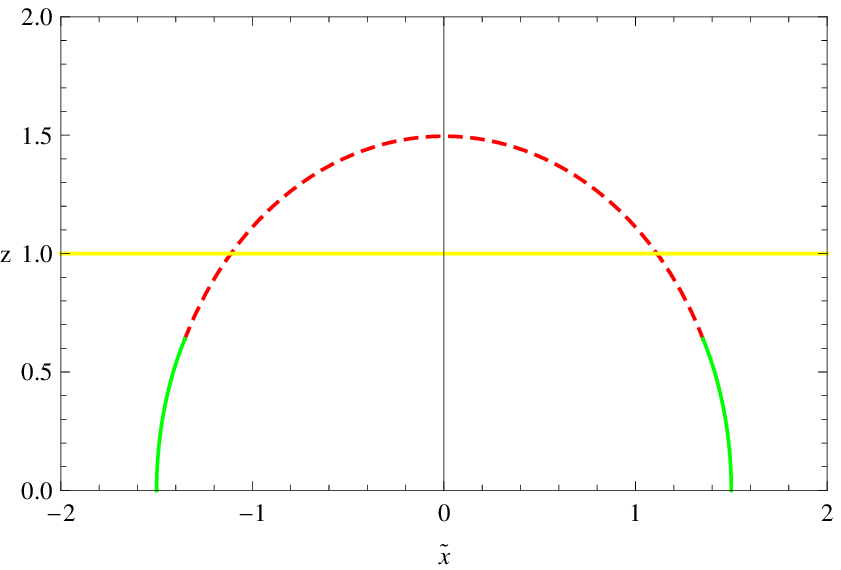}
\label{shell2_RN}
}
\subfigure[$v_*=-0.856, \alpha=0.08$]{
\includegraphics[scale=0.55]{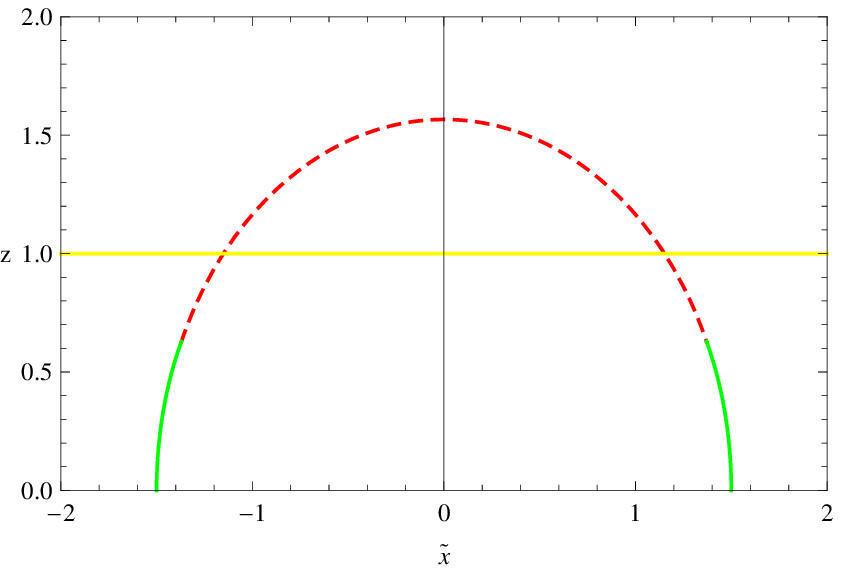}
\label{shell3_RN} }
\subfigure[$v_*=-0.456, \alpha=-0.1$]{
\includegraphics[scale=0.55]{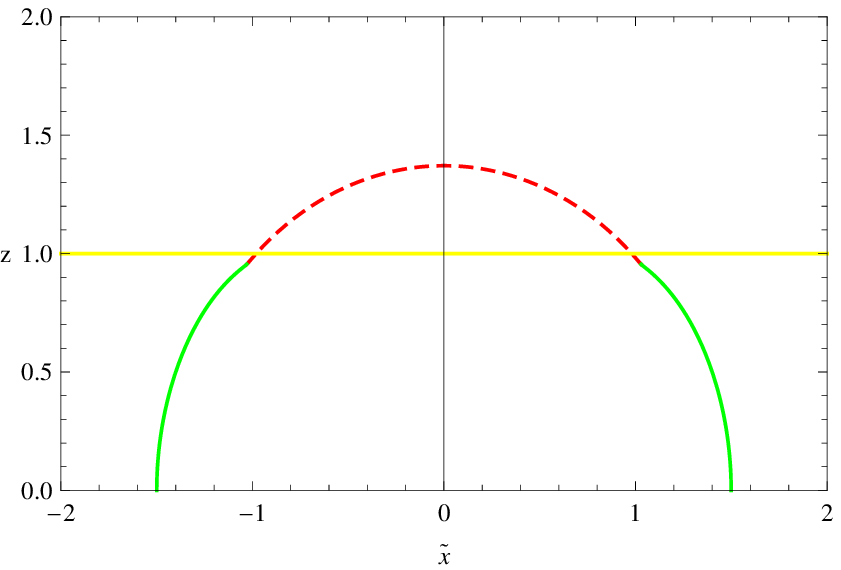}
\label{shell4_RN}
}
\subfigure[$v_*=-0.456, \alpha=0.0001$]{
\includegraphics[scale=0.55]{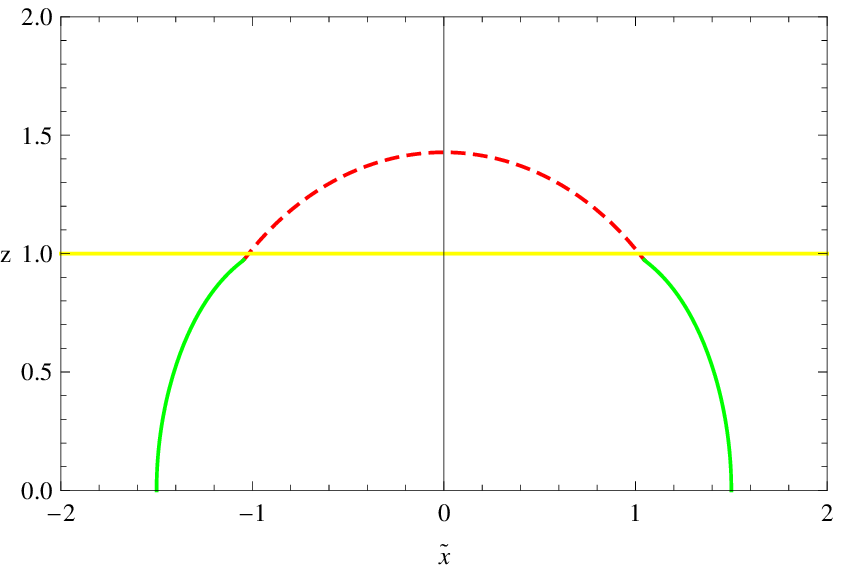}
\label{shell5_RN} }
\subfigure[$v_*=-0.456, \alpha=0.08$]{
\includegraphics[scale=0.55]{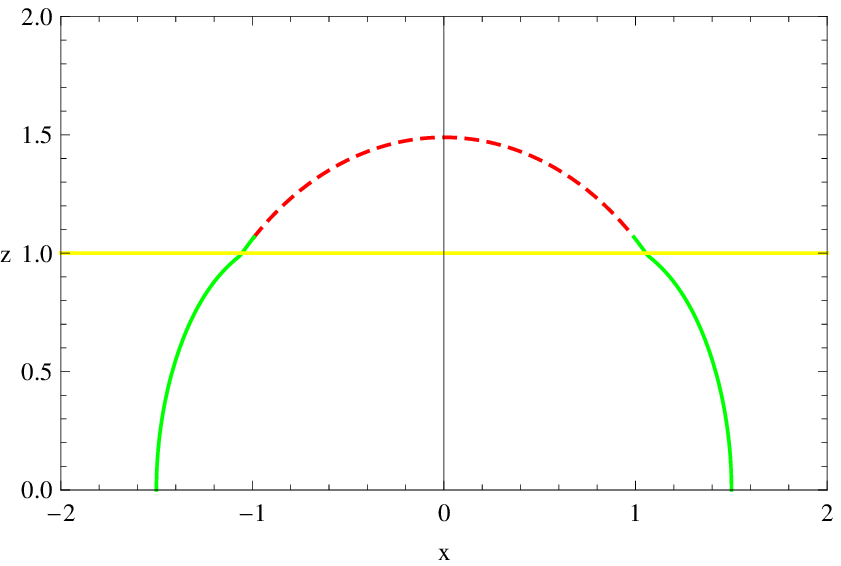}
\label{shell6_RN}
}
\subfigure[$v_*=0.156, \alpha=-0.1$]{
\includegraphics[scale=0.55]{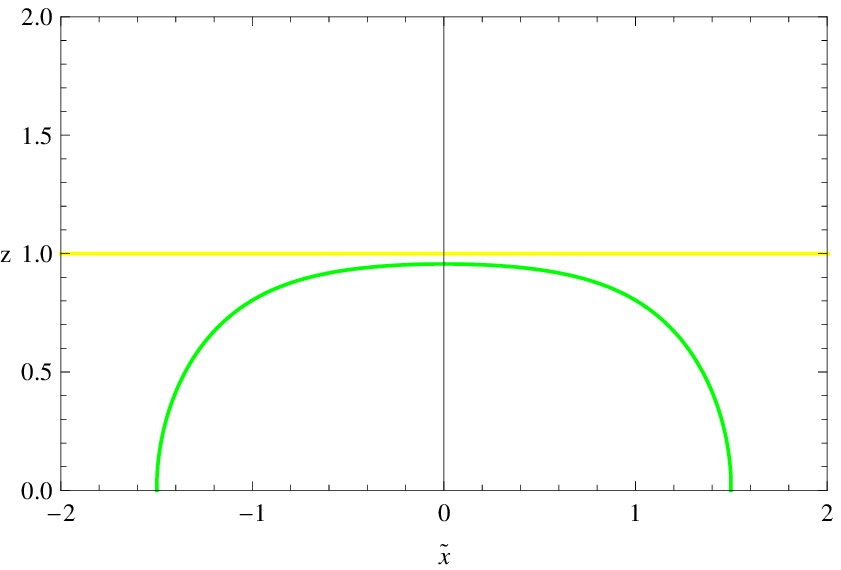}
\label{shell7RN} }
\subfigure[$v_*=0.156, \alpha=0.0001$]{
\includegraphics[scale=0.55]{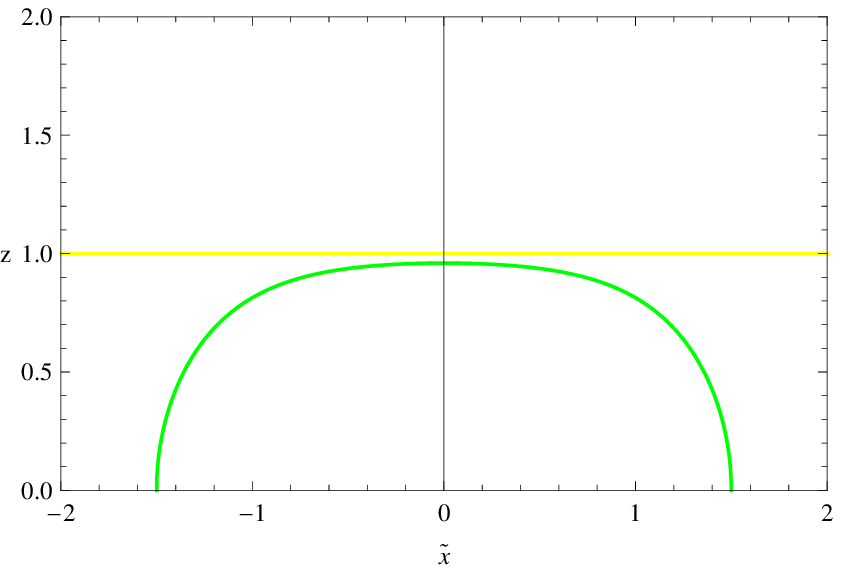}
\label{shell8RN} }
\subfigure[$v_*=0.156, \alpha=0.08$]{
\includegraphics[scale=0.55]{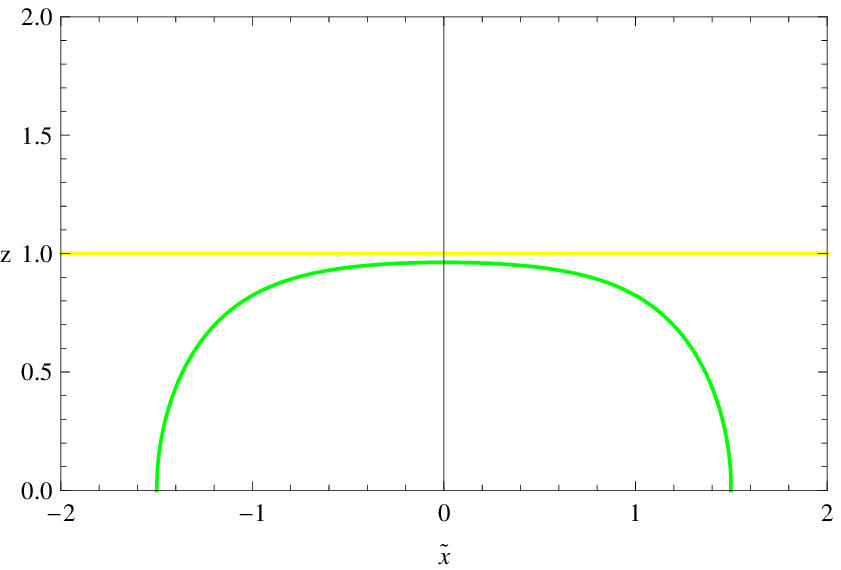}
\label{shell9RN} }
 \caption{\small Motion profile of the geodesics in the Gauss-Bonnet Vaidya AdS black brane. The separation of the
boundary field theory operator pair is $\tilde{\ell}=3$. The black brane
horizon is indicated by the yellow line. The  position of the shell is described by the junction between the dashed red line and the green line.} \label{fig1}
\end{figure}

Firstly let us turn to the  equations of motion of the geodesic in (\ref{gequation}). Using the boundary conditions in (\ref{initial}), we can get the
solutions of $z(x)$ directly for different  $\alpha$. Here, we are interested in the  effect of the Gauss-Bonnet coefficient  $\alpha$ on the  motion
profile of the geodesics.  Considering the  constraint
of
causality of dual field theory on the boundary, we take  $\alpha=-0.1, 0.0001, 0.08$ as examples. The concrete numerical results are shown in
 Figure (\ref{fig1}),  in which  the  horizontal direction is the motion profile of the geodesics for different Gauss-Bonnet coefficients while the
 vertical direction is the motion profile of the geodesics for different initial times.
The  horizontal direction shows  that the Gauss-Bonnet coefficient affects the position of the shell. This phenomenon is most obvious
 for the case $v_*=-0.456$. For  $\alpha=-0.1$, the shell is outside the horizon of the Gauss-Bonnet   AdS  black brane, however as  the Gauss-Bonnet coefficient
increases to  $\alpha=0.08$, the shell drops into the  horizon of the black brane. In other words, for  $\alpha=-0.1$ the  quark
gluon plasma  in the conformal field theory is thermalizing while for  $\alpha=0.08$ it is thermalized.
The thermalization times are listed in Table (\ref{tab:g}). It is shown that as $\alpha$ increases, the thermalization time  decreases for the same
initial time. That is, as the Gauss-Bonnet coefficient
grows larger, the  quark
gluon plasma  is easier to be thermalized.
The vertical direction in  Figure (\ref{fig1}) shows the motion  profile of the geodesics for a fixed  $\alpha$.
As the initial time increases step by step, the shell approaches to the horizon and finally drops into there. In this case,
a static Gauss-Bonnet AdS black  brane  forms  and the thermalization ends up.

 Having the numerical result of $z(x)$, we can study
the  renormalized geodesic length  according to (\ref{lren}). As done in \cite{GS}, we compare $\delta \tilde{L}$
 at each time with the final values $\delta \tilde{L}_{GB}$, obtained in a static Gauss-Bonnet  AdS black brane, \emph{i.e.} $m(\mu)=M$.
  In this case, the thermalized state  is labeled by the zero point of the vertical coordinate in each picture.
 To get an observable quantity that is $\tilde{l}$ independent, we will plot the quantity $\delta L=\delta \tilde{L}/\tilde{l}$.
\begin{table}
\begin{center}\begin{tabular}{l|c|c|c}
 \hline
                    &  $\alpha$=-0.1       &   $\alpha$=0.0001  &   $\alpha$=0.08    \\ \hline
  $v_*$=-0.856     & 0.691683      &  0.625454    &  0.560182    \\ \hline
 $v_*$=-0.456          &1.01534        & 0.949695     & 0.888231      \\ \hline
 $v_*$=0.156           & 1.56525      & 1.50154    &  1.44083      \\ \hline
\end{tabular}
\end{center}
\caption{The thermalization time $t_0$ of the geodesic probe for different Gauss-Bonnet coefficient $\alpha$ and different initial time $v_{\star}$.}\label{tab:g}
\end{table}
\begin{figure}
\centering
\subfigure[$\tilde{l}=2$]{
\includegraphics[scale=0.55]{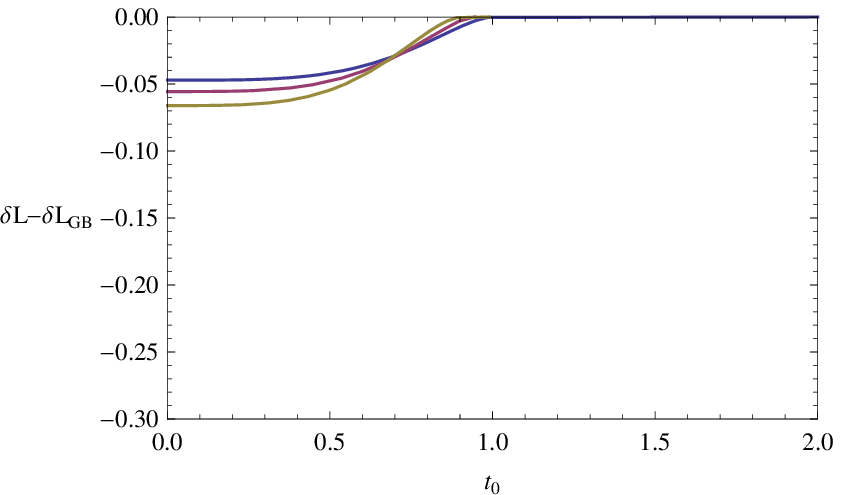}
\label{shell1_RN} }
\subfigure[$\tilde{l}=3$]{
\includegraphics[scale=0.55]{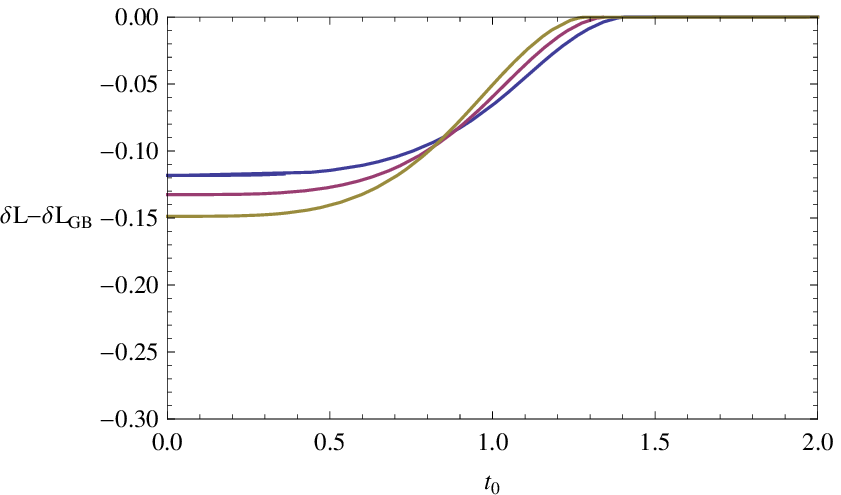}
\label{shell2_RN}
}
\subfigure[$\tilde{l}=4$]{
\includegraphics[scale=0.55]{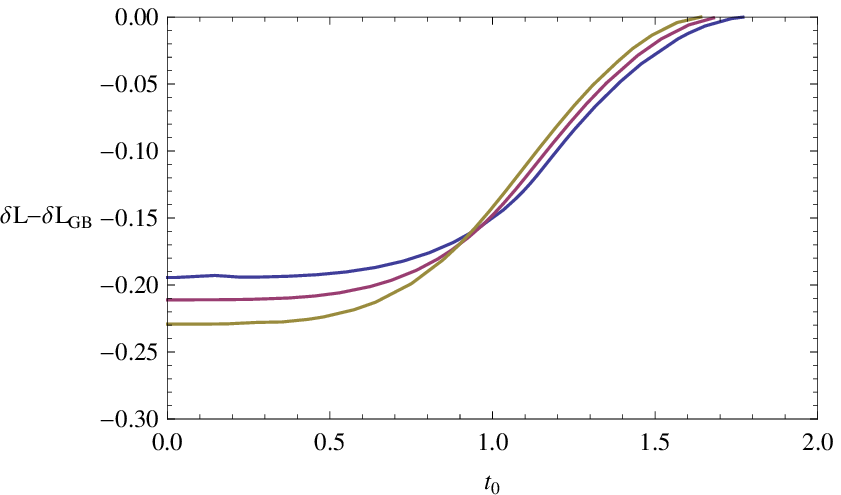}
\label{shell3_RN} }
 \caption{\small Thermalization of the renormalized geodesic lengths in a   Gauss-Bonnet Vaidya AdS  black brane for different Gauss-Bonnet coefficients $\alpha$ and different boundary separations $\tilde{l}$.
 The green line, red line and purple line correspond to  $\alpha=-0.1, 0.0001, 0.08$ respectively.} \label{figa2}
\end{figure}
Figure (\ref{figa2}) gives the relation between the  renormalized geodesic length and thermalization time for different boundary separation that varies horizontally. In each picture, the vertical axis
indicates the renormalized geodesic length while the horizontal axis indicates the time $t_0$.
 From the  same color line, \emph{e.g.}
green line, in (a), (b) and (c) in Figure (\ref{figa2}),  we know that as the separation distance
 of local quantum field theory operators at the boundary increases, the
thermalization time raises for a fixed Gauss-Bonnet coefficients $\alpha$. This result is consistent with that in \cite{Balasubramanian1, Balasubramanian2}, which implies the UV thermalizes first. For a fixed separation distance, \emph{e.g.} $\tilde{l} = 2$,  the thermalization time decreases as
$\alpha$ becomes larger. This phenomenon has been also observed previously when we study the motion profile of the geodesic.  In \cite{GS}, the effect of charge on the thermalization time is investigated, it was shown that there is an enhancement of the thermalization time as the chemical
potential over temperature ratio increases. Obviously, the  Gauss-Bonnet coefficient has an opposite effect on the  renormalized geodesic length compared
with that case. In addition, in Figure (\ref{figa2}), we observe  that for a fixed boundary separation there is always a time range in which  the  renormalized geodesic length takes the same value nearly. That is, during that time range,
the Gauss-Bonnet coefficient has little effect on the   renormalized geodesic length.

\begin{figure}
\centering
\subfigure[$v_*=-0.252, \alpha=-0.1$]{
\includegraphics[scale=0.51]{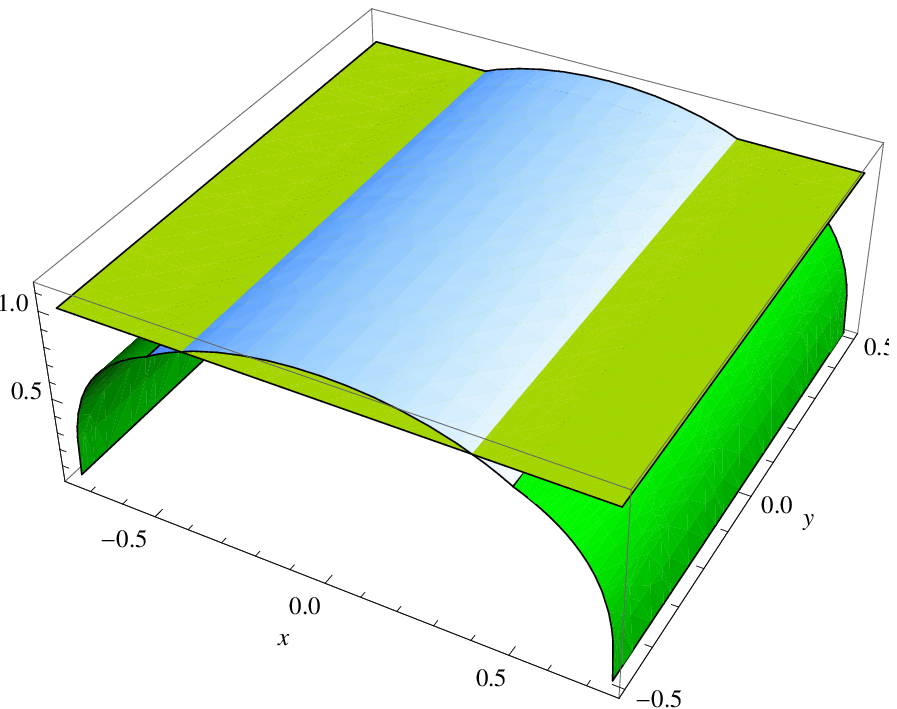}
\label{shell1_RN} }
\subfigure[$v_*=-0.252, \alpha=0.0001$]{
\includegraphics[scale=0.51]{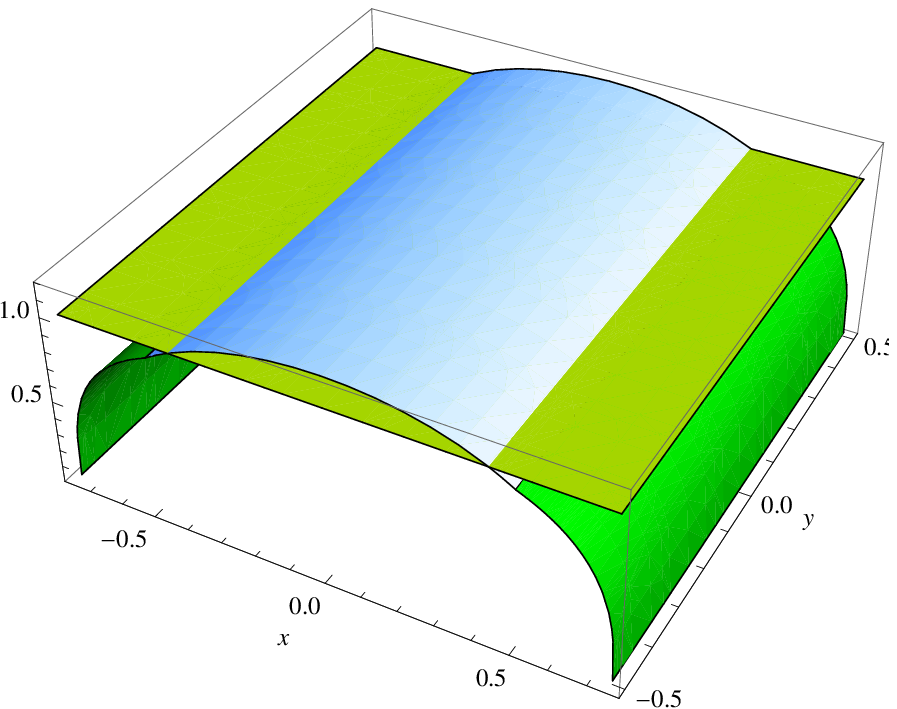}
\label{shell2_RN}
}
\subfigure[$v_*=-0.252, \alpha=0.08$]{
\includegraphics[scale=0.51]{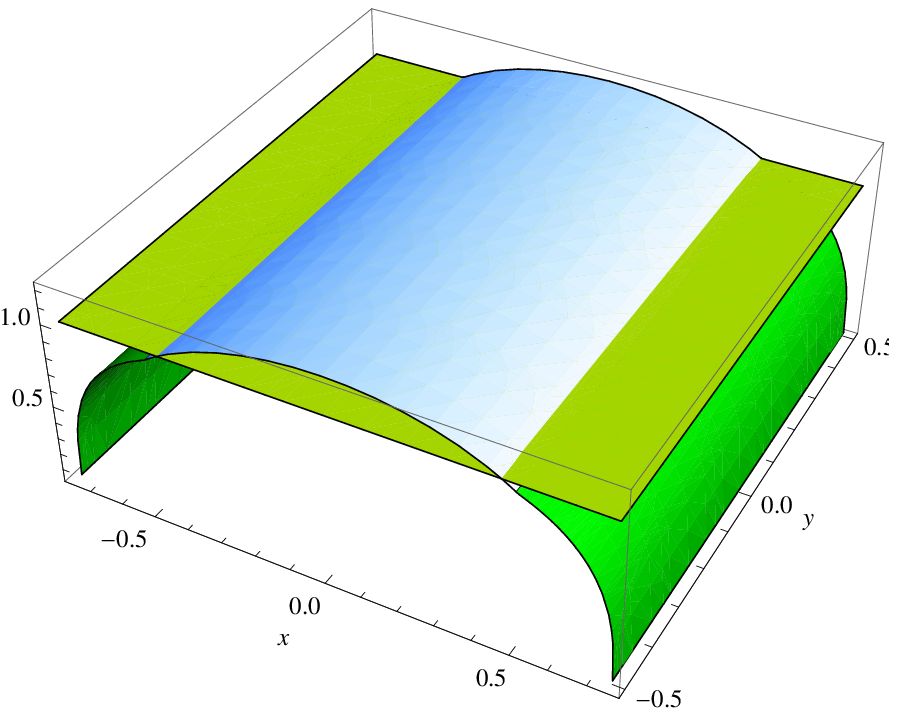}
\label{shell3_RN} }
 \caption{\small Motion profile of the minimal area   in the  Gauss-Bonnet Vaidya AdS black brane. The boundary separation along the $x$ direction is $1.5$, and along the $y$ direction is $1$. The yellow surface is the location of the horizon.  The  position of the shell is is described by the junction between the  white surface and the green surface.} \label{figa11}
\end{figure}

\begin{figure}
\centering
\subfigure[$\tilde{l}=1$]{
\includegraphics[scale=0.55]{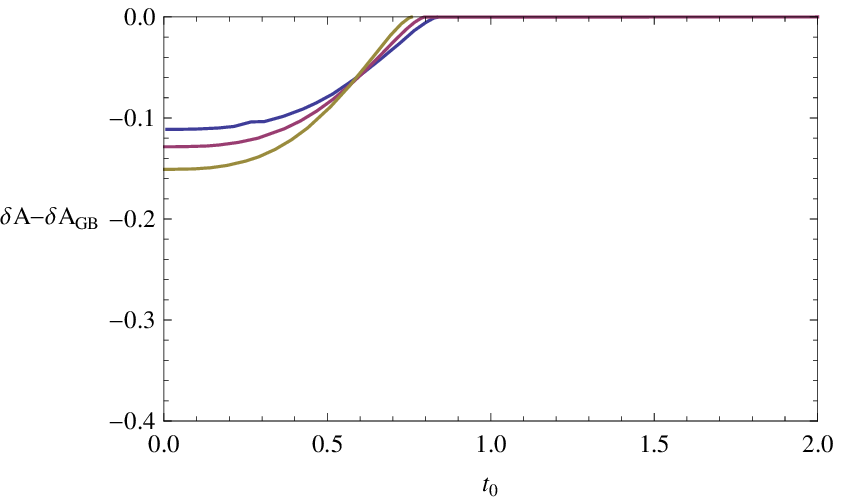}
\label{shell1_RN} }
\subfigure[$\tilde{l}=1.5$]{
\includegraphics[scale=0.55]{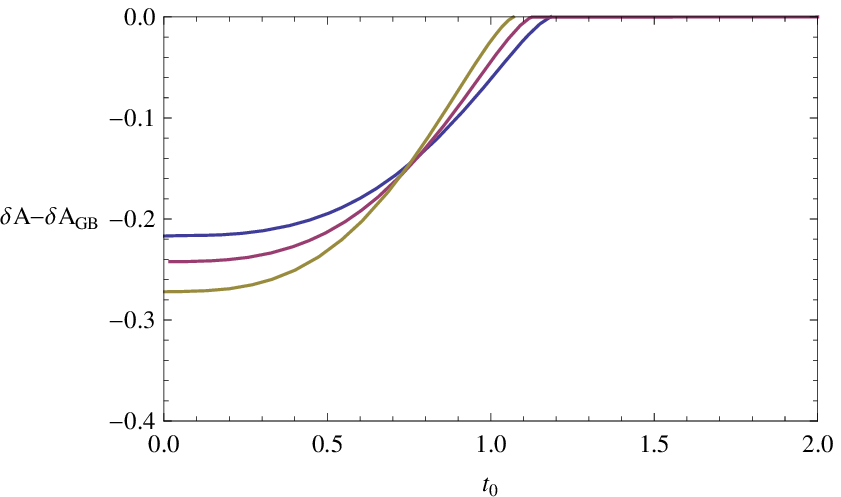}
\label{shell2_RN}
}
\subfigure[$\tilde{l}=2$]{
\includegraphics[scale=0.55]{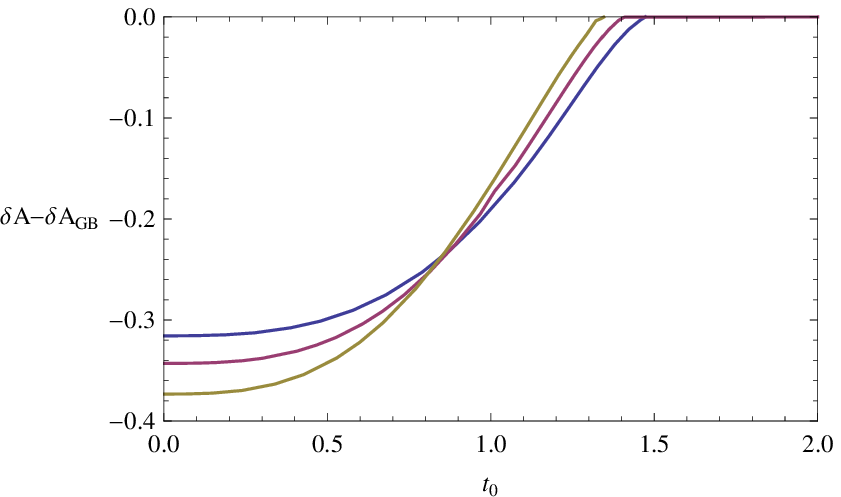}
\label{shell3_RN} }
 \caption{\small Thermalization of the renormalized minimal surface  area  in a  Gauss-Bonnet Vaidya AdS black brane  for different Gauss-Bonnet coefficients $\alpha$ and different boundary separations $\tilde{l}$.
 The green line, red line and purple line correspond  to  $\alpha=-0.1, 0.0001, 0.08$ respectively.} \label{figa33}
\end{figure}

Adopting similar strategy, we also can study the motion profile of minimal area   and the relation between the renormalized minimal  area surface  and thermalization
 time. Based on the motion equations in (\ref{aequation}) and the boundary conditions in  (\ref{initial}),
the numerical solution of $z(x)$  can be produced. In this case, we can get the  motion profile of minimal area  for different $\alpha$,
which are shown in Figure (\ref{figa11}). From  this figure, we know that as the Gauss-Bonnet coefficient increases, the shell surface approaches
 to the horizon surface step by step.
 The thermalization time for different $\alpha$ have been listed in Table (\ref{taba1}). It is obvious that the thermalization time decreases
 as $\alpha$ becomes larger. This behavior is similar to that of the geodesics. In addition, we also can substitute the numerical result
  of $z(x)$ into  (\ref{aren}) to get the renormalized minimal  area surface. Similar to the case of geodesic, we will plot $\delta A-\delta A_{GB}$,
   where  $\delta A=\delta \tilde{A}/\tilde{l}$ and $\delta A_{GB}$ is the  renormalized minimal  area surface    for a  static  Gauss-Bonnet AdS black brane.
   The relation between the renormalized minimal  area surface    and thermalization time
 is given in Figure  (\ref{figa33}), in which the vertical axis indicates the renormalized minimal  area surface while the horizontal axis indicates
 the thermalization time $t_0$. We find that  as the separation distance increases, the
thermalization time raises for a fixed Gauss-Bonnet coefficient, which confirms the fact that the UV modes thermalize first. And for a fixed separation distance, the thermalization time decreases as
$\alpha$ becomes larger. That is to say,  the larger the Gauss-Bonnet coefficient is, the easier the quark gluon plasma  thermalizes. This behavior
is similar to that of the geodesic which is given in Figure (\ref{figa2}). As the case of the renormalized geodesic length, we find that
in Figure  (\ref{figa33}), there is also an overlapped region, where the Gauss-Bonnet coefficient has little effect on the renormalized minimal area surface.

~

\begin{table}
\begin{center}\begin{tabular}{l|c|c|c}
 \hline
                    &  $\alpha$=-0.1       &   $\alpha$=0.0001  &   $\alpha$=0.08    \\ \hline
  $v_*$=-0.252    & 1.01732      &  0.963057   & 0.911427   \\ \hline
\end{tabular}
\end{center}
\caption{The thermalization time $t_0$ of the minimal area   probe  for different Gauss-Bonnet coefficients $\alpha$ with the same  initial time $v_{\star}$.}\label{taba1}
\end{table}



~

\section{Conclusions}
The  thermalization time scale of the dual boundary field theory in  Gauss-Bonnet gravity is studied in the framework of  AdS/CFT correspondence.
 The thermalization process in the dual field theory is modeled by the collapsing of a shell of dust that  interpolates between a pure AdS and a
  Gauss-Bonnet AdS black brane, in which the ground state is sufficiently excited by the injection of energy and followed by the thermalization.
   The two-point functions and expectation values of Wilson loops are chosen as the thermalization probes, which
    are dual to the  renormalized geodesic length and minimal  area surface   in the bulk. The effect of the  Gauss-Bonnet coefficients
   on the thermalization time is studied.
We  first obtain the motion profiles of the geodesic and minimal surface   and find that for both cases the thermalization time
decreases as the Gauss-Bonnet coefficient increases. We reproduce this result by studying the relation between the  renormalized
geodesic length and time as well as the renormalized minimal surface  area and time respectively. In addition, for both the thermalization probes,
 we observe an overlapped region where the
Gauss-Bonnet coefficient has few influence on them for a fixed boundary separation. The reason for this phenomenon maybe arises
 from the delay of the thermalization. As stressed in \cite{Balasubramanian2}, the thermalization only becomes fully
apparent at distances of the order of the thermal screening length $\tilde{l}_D\sim(\pi T)^{-1}$, where $T$ is the temperature of
 the dual conformal field. From (\ref{temperature}), we know that the temperature is independent of the Gauss-Bonnet coefficient,
  so we can conclude safely that the thermalization for different $\alpha$ begins apparently at the almost same distance, leading to an overlap.

On the other hand, it is known  from the viewpoint of holography that IIB string theory on AdS5 $\times S^5$ background is dual to $D=4, ~N=4, ~SU(N_c) $ super Yang-Mills theory, and higher
derivative corrections from string correspond
to finite 't Hooft coupling corrections. The most common corrections to the string are  described by the terms $R^4$ and $R^2$ \cite{1008.2430}. Recently Ref. \cite{1305.2237} investigated the effect of  $R^4$ corrections on the thermalization time scale in the framework of  supergravity. They found that  finite 't Hooft coupling corrections decrease  just a very little the thermalization
time of UV modes, while they produce the opposite trend for IR modes. In this paper, we investigated one case of  $R^2$ corrections, namely the Gauss-Bonnet gravity. We found that the Gauss-Bonnet coefficient will enhance the thermalization time  for both the thermalization probes. It is also interesting to study the effect of $R^2$ corrections to the thermalization time scale in the framework of  supergravity to compare with our work and that in \cite{1305.2237}.

\section*{Acknowledgements}

 Xiao-Xiong Zeng would like to thank Hongbao Zhang for his encouragement and various valuable suggestions during this work. He is also grateful to
Dami$\acute{a}$n Galante, Weijia Li, Kai Lin, and Wieland Staessens for helpful discussions on numerics.
 This work is supported in part by the National Natural Science Foundation of China (Grant Nos.10773002, 10875012, 11175019) and the Fundamental Research Funds for the Central Universities under Grant No.105116.

\end{document}